# FRONTIER SCIENCE IN A QUANTUM EXPERIMENT: AEgIS AT CERN

Rozwój nowego obszaru nauki w eksperymencie kwantowym: AEgIS w CERN

According to the best understanding of the laws governing our Universe, matter and antimatter were produced in equal amounts at the onset of the Big Bang. For reasons that are presently unknown, only matter emerged as the prevailing constituent, and today, only minor quantities of antimatter can be located. This dilemma stands as one of the pivotal unresolved questions in modern physics. Precision studies of exotic systems containing both might be key to understanding their differences better.

Cold antimatter experiments conducted approximately 30 years ago at CERN primarily served as exploratory measurements [1]. During that period, the techniques available needed to be revised to enable researchers to deliberately generate exotic atoms in a controlled manner. Their production was attained through bulk materials. At present, our endeavour involves surpassing these initial attempts [2,3]. This is achieved by amalgamating techniques derived from particle, nuclear, plasma, and atomic physics. The ultimate goal is to achieve a controlled synthesis of these atoms directly within traps under ultra-high vacuum conditions that are ready for further experimentation [4].

Ph.D., D.Sc. Georgy Kornakov, M.Sc. Jakub Zieliński,
Warsaw University of Technology, Warsaw, Poland

**ABSTRACT**

Precise manipulation of matter at the atomic or molecular level has provided the path for the nanotechnological revolution impacting diverse fields such as biology, medicine, material science, quantum technologies, and electronics. At the Antiproton Decelerator facility at CERN, the AEgIS experiment utilises state-of-the-art technology to store and manipulate synthesised exotic atoms containing both matter and antimatter. Such experiments lay the groundwork for a better understanding of the fundamental interactions and hold the potential to unravel the enigma of the absence of antimatter in our universe. Additionally, the developed techniques advance the technological frontier of controlling the quantum states of ions, a critical aspect of quantum sensing and quantum computing applications.

**STRESZCZENIE**

Precyzyjna manipulacja materią na poziomie atomowym lub molekularnym otworzyła drogę do rewolucji nanotechnologicznej, która ma wpływ na różne dziedziny, takie jak biologia, medycyna, materiałoznawstwo, technologie kwantowe czy elektronika. W obiekcie Antiproton Decelerator w CERN eksperyment AEgIS wykorzystuje najnowocześniejszą technologię do przechowywania i manipulowania zsyntetyzowanymi egzotycznymi atomami zawierającymi zarówno materię, jak i antymaterię. Takie eksperymenty kładą podwaliny pod lepsze zrozumienie podstawowych oddziaływań i mają potencjał do rozwiązania zagadki braku antymaterii w naszym wszechświecie. Ponadto opracowane techniki przesuwają granicę technologiczną w zakresie kontrolowania stanów kwantowych jonów, krytycznego aspektu kwantowych zastosowań wykrywania i obliczeń kwantowych.

**SŁOWA KLUCZOWE:** pułapki jonowe, atomy Rydberga, antymateria, CERN, antywodór, zasada słabej równoważności

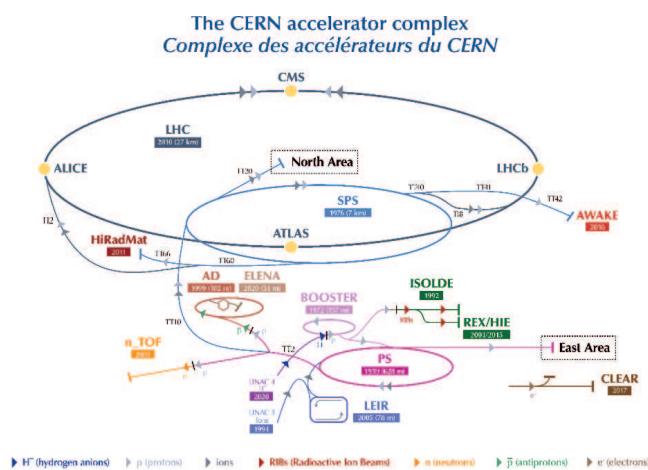

**Fig. 1.** The CERN accelerator complex. LHC is in dark blue. The AD and ELENA are shown among the smaller machines, which slow down antiprotons [CERN-GRAPHICS-2019-002, Credits for the graph: Esma Mobs, 2019]
**Rys. 1.** Kompleks akceleratorów CERN. LHC jest w kolorze ciemnoniebieskim. Wśród mniejszych maszyn znajdziemy AD i ELENA, które służą do spowalniania antyprotonów, [CERN-GRAPHICS-2019-002, Credits for the graph: Esma Mobs, 2019]

To explore the attributes of antimatter, the Antimatter Factory has been constructed at the CERN Antiproton Decelerator facility [5,6]. The CERN accelerator complex is shown in *Figure 1*. It is positioned as the only abundant source of low-energy baryonic antimatter. The AEgIS (Antimatter Experiment, gravity, interferometry, and spectroscopy) experiment [7] is concurrently pursuing the synthesis of exotic atoms containing antimatter, along with antihydrogen atoms, for detailed characterisation of their properties and their coupling to the Earth's gravitational field.

The experimental examination of these distinctive atoms grants access to the thresholds of nuclear and electromagnetic forces. The interactions and equivalences of fundamental physics laws can be meticulously scrutinised, comparing the experimental results with state-of-the-art theories. Differences among them might point to the existence of new physics.

## EXPERIMENT AND INFRASTRUCTURE

CERN has a rich tradition of delving into fundamental physics across extremely low and high-energy scales. In 2011 it was approved the construction of the Extra Low ENergy Antiproton (ELENA) ring [6]. The primary purpose of this infrastructure is to facilitate the exploration of phenomena not described by the Standard Model or General Relativity. The facility enables precise and detailed measurements of antihydrogen and low-energy antiprotons. This is the world's single source of low-energy antiprotons.

Technically, the antiprotons are produced in collisions of relativistic protons with an iridium target. The protons are accelerated by the Proton Synchrotron to an energy of approximately 26 GeV and afterwards extracted to the metal target. Then, bending and focusing magnets select antiprotons from the produced particles in the collisions. These antiprotons are transferred to a synchrotron called the Antiproton Decelerator [5] which is meant to diminish their energy (and the temperature). The antiparticles are slowed down using techniques such as electron and stochastic cooling. After several cycles, the antiprotons have reduced their energy to 5.3 MeV. Then, they are transferred to the next decelerator ring for further cooling. ELENA is a small 30 m circumference synchrotron to further decelerate antiprotons from the AD from 5.3 MeV down to 100 keV and finally to transfer them to the experiments present at the antimatter laboratory.

One of the experiments which receives the antiprotons is AEgIS. Its apparatus implements two cylindrical cryostats containing two superconducting magnets of 5 T and 1 T, respectively. A series of cylindrical electrodes inside each magnet forms a Malmberg-Penning trap arrangement and allows radial and axial confinement of charged particles. The antiprotons are trapped and cooled to a few K by sympathetic cooling with an electron cloud previously stored inside the trap. The antiprotons are then ballistically transferred from the 5 T trap to the 1 T antihydrogen production region, where they are re-caught in flight. Rydberg-excited positronium atoms, a bound state of an electron and a positron, are directed towards the antiproton plasma to synthesise a pulsed beam of Rydberg antihydrogen by a resonant charge-exchange reaction [8]. The 1 T trap is shown in *Figure 2*. The trapped matter and antimatter manipulation is done via lasers, which can also be used to perform spectroscopy.

> THE WUT GROUP HAS BEEN A MEMBER OF THE AEgIS COLLABORATION SINCE 2020.

## CERN, AEgIS COLLABORATION AND THE AEgIS-PL CONSORTIUM

The WUT group has been a member of the AEgIS Collaboration since 2020. In September 2021, IFPAN, UMK and PW signed the AEgIS-PL Consortium, which provides the framework for the joint effort within the AEgIS experiment of the three institutions. The consortium members offer the collaborative project different know-how and skills gained in their respective fields.

WUT contributes to the experiment with the following:
- Detailed knowledge of simulations in GEANT and modelling of nuclear interactions.
- Engineering of electronics and control systems with sub-ns time synchronisation compatible with the Sinara/ARTIQ open hardware and software environments. (Co-Investigator)
- Detailed knowledge of design principles of the experiment and preparation of the measurement.

UMK contributes with:
- Experimental knowledge of ion traps.
- Expertise in ultra-precise spectroscopy techniques such as laser spectroscopy.
- Experience in preparing ultra-cold atoms.
- Reference frequency and time.

IFPAN contributes with:
- Theoretical expertise of exotic atomic systems and QED.

CERN/AEgIS collaboration provides:
- Access to a high-intensity source of low-energy antiprotons.
- Access to ultra-high vacuum, superconducting magnets, atomic traps, lasers, detectors and control equipment.

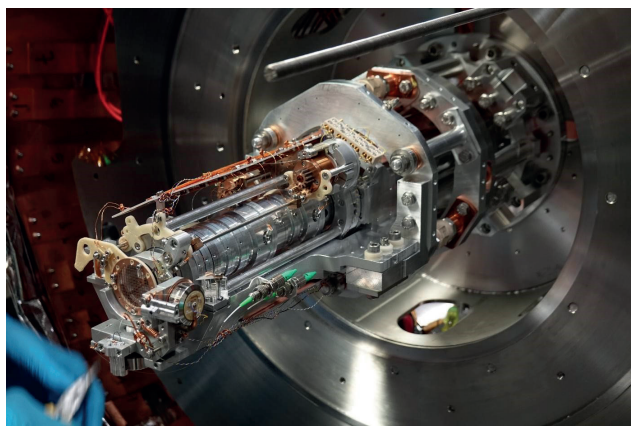

**Fig. 2.** One of the AEgIS Penning-Malmberg traps is used for the synthesis of antihydrogen
**Rys. 2.** Jedna z pułapek AEgIS Penninga-Malmberga stosowana do syntezy antywodoru

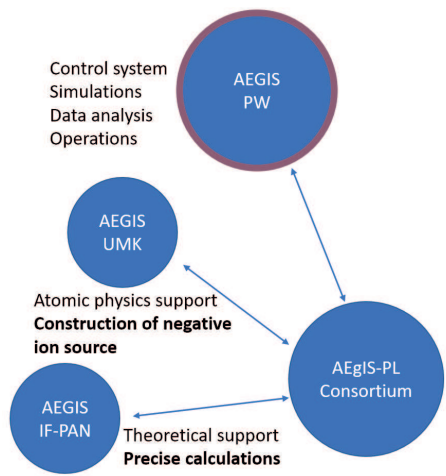

**Fig. 3.** Structure of the AEgIS-PL Consortium
**Rys. 3.** Struktura Konsorcjum AEgIS-PL

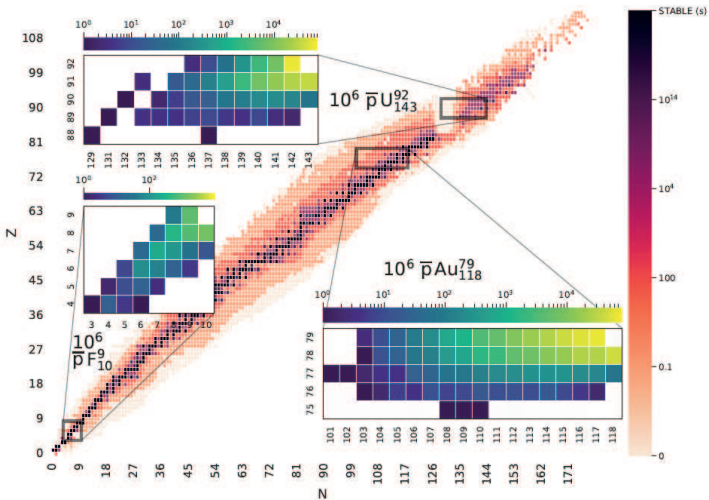

**Fig. 4.** Populations of Highly Charged Ions obtained in annihilations of antiprotons on cold atoms of fluorine, gold and uranium [4]
**Rys. 4.** Populacje silnie naładowanych jonów otrzymane w anihilacjach antyprotonów na zimnych atomach fluoru, złota i uranu [4]

## SUMMARY

Advancements in ion trap technologies in the last decade have brought the possibility of exploring exotic states of matter and antimatter by using precision spectroscopy techniques or interferometers. The AEgIS experiment has recently achieved pulsed production of Rydberg antihydrogen. It is also taking the first steps towards synthesising cold antiprotonic atoms, which could be gateways to various systems for QCD, QED and physics tests beyond the Standard Model of particle physics. ❖

### ACKNOWLEDGEMENTS

*This work is funded by the Research University – Excellence Initiative of Warsaw University of Technology via the strategic funds of the Priority Research Center of High Energy Physics and Experimental Techniques and by the Polish National Science Centre under agreements no. 2022/45/B/ST2/02029, and no. 2022/46/E/ST2/00255 and by the Polish Ministry of Education and Science under contract no. 2022/WK/06.*

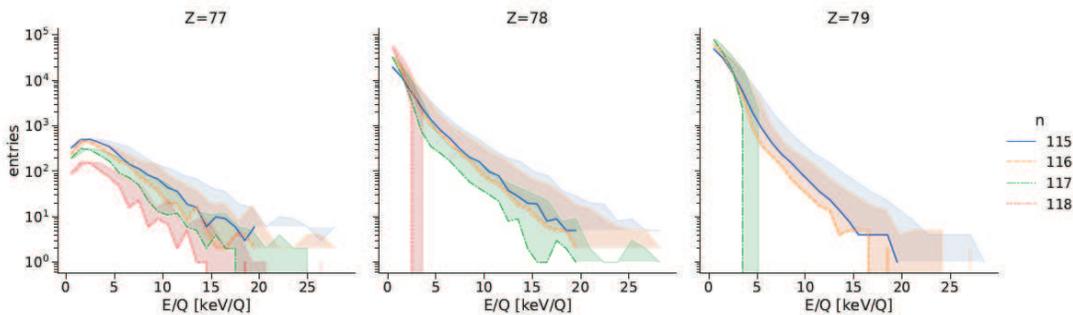

**Fig.5.** Energy spectra of produced ions from capturing antiprotons by gold atoms [4]
**Rys. 5.** Widma energetyczne wytworzonych jonów z wychwytu antyprotonów przez atomy złota [4].

## PRODUCTION OF HIGHLY CHARGED IONS IN TRAPS

The methodology developed for antihydrogen can also be adapted to other exotic atoms containing matter and antimatter simultaneously. The scheme has been studied using detailed Monte Carlo simulations using the GEANT4 framework [4]. As depicted in *Figure 4*, three examples of end products are shown after the annihilation of antiprotons on the surface of the distinct nuclei. The simulation suggests that medium-heavy and heavy initial atoms are suitable for efficiently producing entirely or almost wholly stripped ions directly in traps. The energy of the produced fragments is shown in *Figure 5*.

Ions characterised by energies over charge falling below a few tens of keV can be re-captured immediately after production using either nested trap configurations or pulsed electrodes. Further reduction of the energy of the fragments can be attained by mixing the plasmas with positive particles like protons or positrons using sympathetic cooling, allowing them to perform precision studies immediately after their production. Once the populations of ions are again cold, they are ready for further experimentation or spectroscopic measurements, even of short-lived radioactive sources.